\documentclass[longnamesfirst,apj]{emulateapj}
\usepackage{apjfonts,natbib,epsfig}
\tighten

\newcommand{\beq}{\begin{equation}}
\newcommand{\eeq}{\end{equation}}
\newcommand{\bea}{\begin{eqnarray}}
\newcommand{\eea}{\end{eqnarray}}

\begin{document}
\title{The Cosmic Ray Precursor of Relativistic Collisionless Shocks: A Missing Link in Gamma-Ray Burst Afterglows}
\author{Milo\v s Milosavljevi\'c\altaffilmark{1,2} and Ehud Nakar\altaffilmark{1}}
\altaffiltext{1}{Theoretical Astrophysics, Mail Code 130-33, California Institute of Technology, 1200 East California Boulevard, Pasadena, CA 91125.  }
\altaffiltext{2}{Hubble Fellow.}
\righthead{Cosmic Ray Precursor in GRB Shocks}
\lefthead{MILOSAVLJEVI\'C \& NAKAR}

\begin{abstract}

Collisionless shocks are commonly argued to be the sites of cosmic ray (CR) acceleration.  We study the influence of CRs on weakly magnetized relativistic collisionless shocks and apply our results to external shocks in gamma-ray burst (GRB) afterglows.  The common view is that the transverse Weibel instability (TWI) generates a small-scale magnetic field that facilitates collisional coupling and thermalization in the shock transition.  The TWI field is expected to decay rapidly, over a finite number of proton plasma skin depths from the transition.  However, the synchrotron emission in GRB afterglows suggests that a strong and persistent magnetic field is present in the plasma that crosses the shock; the origin of this field is a key open question.  Here we suggest that the common picture involving TWI demands revision.  Namely, the CRs drive turbulence in the shock upstream on scales much larger than the skin depth.  This turbulence generates a large-scale magnetic field that quenches TWI and produces a magnetized shock. The new field efficiently confines CRs and enhances the acceleration efficiency. The CRs modify the shocks in GRB afterglows at least while they remain relativistic. The origin of the magnetic field that gives rise to the synchrotron emission is plausibly in the CR-driven turbulence.  We do not expect  ultrahigh energy cosmic ray production in external GRB shocks.  


\end{abstract}

\section{Introduction}

\setcounter{footnote}{0}

Collisionless shocks are observed on all astrophysical scales. The diffusive shock acceleration (DSA) mechanism is believed to accelerate cosmic rays (CRs) in these shocks \citep{Bell:78,Blandford:78,Blandford:87}.  The CRs 
can carry a substantial fraction of the energy of the shock and thus 
the CR pressure can influence the structure of the shock \citep{Eichler:79,Blandford:80,Drury:81,Ellison:84}. 
This picture has recently  
received support from  X-ray observations of supernova (SN) remnants \citep{Warren:05}. Recently,
\citet{Bell:04,Bell:05} has shown that CRs in SNe can drive turbulence and amplify magnetic fields in the shock upstream, and \citet{Dar:05} speculate that relativistic jets could do the same.

The current lore is that weakly magnetized relativistic
collisionless shocks are mediated by the transverse Weibel instability
(TWI; \citealt{Weibel:59,Fried:59}). TWI produces a
magnetic field near equipartition that provides collisional coupling in the shock transition layer 
\citep{GruzinovWaxman:99,Medvedev:99}. TWI is universally
observed in two- and three-dimensional particle-in-cell (PIC)
simulations of unmagnetized colliding shells
(e.g.,~\citealt{Lee:73,Gruzinov:01,Silva:03,
Frederiksen:04,Jaroschek:04,Nishikawa:05a,Medvedev:05,Kato:05}). 
The resulting field is small-scale, of the order of the proton plasma
skin depth $\lambda_{\rm s}$. 

The role of TWI in particle acceleration and in
the downstream thermodynamics is controversial.
The field is expected to decay rapidly, within a few $\lambda_{\rm s}$ from
the shock transition \citep{Gruzinov:01,Milosavljevic:06}. This
decay is evident in three dimensional PIC simulations \citep{Spitkovsky:05}.
Although there are claims that the decay
saturates at distances of $\sim 100\lambda_{\rm s}$ from the shock
\citep{Silva:03,Medvedev:05}, survival of the 
field over larger distances has not been demonstrated.

According to the popular model of gamma-ray bursts 
(GRBs; e.g., \citealt[and references therein]{Piran:05a}), the
afterglow originates in a relativistic blast wave that propagates
into an ambient $e^-p$ plasma. The afterglow 
emission is ascribed to synchrotron
radiation from nonthermal electrons that gyrate in the shock-generated 
magnetic field.
Detailed studies of GRB spectra and light curves
\citep{Panaitescu:02,Yost:03} have shown that the magnetic energy
density in the emitting region (the downstream shocked plasma) is a
fraction $\epsilon_B\sim 10^{-2}$ to $10^{-3}$ of the internal energy
density.\footnote{Under some circumstances $\epsilon_B$ as low as $\sim10^{-5}$ can fit the data \citep{Eichler:05}.} This field must persist over 
at least a few percent of the width of the blast wave \citep{Rossi:03}.
Its origin has remained a key open question. 

Compressional amplification of the
weak pre-existing magnetic field of the interstellar medium (ISM) yields
$\epsilon_B\sim 10^{-9}$ \citep{Gruzinov:01} and does not
explain the field in the emitting region. 
If TWI does develop in the transition layer, it generates a strong field in the vicinity of the shock transition. 
However, as discussed above, there is no evidence that it can persist
over the required $\sim10^9\lambda_{\rm s}$ from the shock
(e.g.,~\citealt{Piran:05b} and references therein).

X-ray observations of GRB afterglows are modeled as an
optically thin synchrotron spectrum requiring nonthermal
electrons with Lorentz factors as high as $\sim 10^6$. The
observations indicate that electrons, and thus protons as well, are
efficiently accelerated in the  shock
to produce a hard power-law spectrum. DSA can achieve such
Lorentz factors if the circum-burst medium is magnetized at
$\sim 1\ \mu\textrm{G}$ level, as expected in the ISM. 
The weak magnetic field, however,  does not
directly affect the structure of the shock 
(e.g., TWI can still develop in the shock transition).

It seems that a self-consistent picture of relativistic astrophysical
shocks includes high energy particles (CRs) and at least a
weakly magnetized upstream. Here we explore the interaction between these components.
In \S~\ref{sec:amplification}, 
we derive the conditions under which
the CRs drive turbulence in the upstream;
this turbulence generates a large-scale
magnetic field that increases the acceleration
efficiency.\footnote{\citet{Gruzinov:01} and \citet{Waxman:04} speculate that CRs could amplify large-scale magnetic fields in external GRB shocks.} In \S~\ref{sec:grb}, we argue that this mechanism is a candidate solution of the origin of the 
field inferred from
the afterglow emission in external GRB shocks. In \S~\ref{sec:discussion}, we
discuss implications for TWI and for the production of
ultrahigh energy cosmic rays (UHECRs).

\section{Cosmic Ray -- Magnetic Field Interaction}
\label{sec:amplification}

\subsection{Cosmic Ray Trajectories and the Return Current}
\label{sec:trajectories}

Consider a relativistic shock wave with Lorentz factor $\Gamma$ that
accelerates particles to Lorentz factors
$\gamma\gg \Gamma$. 
The accelerated particles are roughly isotropic
in the downstream frame; a fraction of them are moving 
ahead of the shock.
In the upstream frame, the velocities of the accelerated particles
are directed within an
angle $\theta \sim \sqrt{2}/\Gamma$ of the shock normal.\footnote{Unless otherwise
noted, quantities are evaluated in the upstream frame.} If the upstream
contains a magnetic field ${\bf B}$, the field deflects charged
particles moving ahead of the shock.  When the velocity of a
particle is deflected to an angle $\sim 1/\Gamma$ from the shock
normal, the shock catches up and re-absorbs the particle.

The orbit of a particle in the upstream field can be
quasi-circular or diffusive, depending on how the radial
distance $X$ the particle traverses ahead of the shock compares to the relevant magnetic correlation length $\lambda$.  The typical angle
by which the magnetic field deflects the particle is $\Delta\theta
\sim X/r_{\rm g}$ when the orbit is quasi-circular and $\Delta\theta
\sim (\lambda X)^{1/2}/r_{\rm g}$ when the orbit is diffusive, where
\beq
\label{eq:rg}
r_{\rm g}=\frac{\gamma mc^2}{eB(\lambda)}
\eeq 
is the gyroradius of the
particle expressed in terms of the magnetic field $B(\lambda)$ on
scales $\lambda$.  Reabsorption in the shock occurs when $\Delta\theta\sim 1/\Gamma$, and thus the distance traversed by the
particle in the quasi-circular and diffusive regime is
\beq
\label{eq:X}
X\sim \cases{ 
r_{\rm g}/\Gamma , & $(X\lesssim \lambda)$ , \cr
(r_{\rm g}/\Gamma)^2/\lambda , & $(X\gg \lambda)$ .} 
\eeq

In spherical blast waves, 
a particle can be accelerated by repeated shock crossing if the time
separating its emission and re-absorption in the shock $\sim X/c$ is
smaller than the expansion time of the shock  $\sim R/c$, where
$R$ is the shock radius.\footnote{It is also implicitly assumed that some fraction of accelerated particles can diffuse back to the shock after they find themselves in the downstream.} Therefore $X\leq R$ is required for
acceleration. The maximum distance
that the particle reaches from the moving shock transition is
substantially smaller, 
\beq
\label{eq:Delta}
\Delta \sim \frac{X}{\Gamma^2} ,
\eeq
because the particle
and the shock are both moving at about the speed of light.  
Another requirement for acceleration is that the cooling
time be longer than the acceleration time (see \S~\ref{sec:grb}).

In general, protons and electrons are both emitted into the shock
upstream.  However, the synchrotron and self-Compton cooling
inhibit the acceleration of electrons, and so the protons, which we refer to as the CR precursor of the shock, 
can reach higher energies and farther from the shock into the upstream. 
The upstream fluid sees a net positive CR current 
$|{\bf J}_{\rm CR}|\sim e v_{\rm s} n_{\rm CR}$ in the region where it overlaps with the precursor.  Here $v_{\rm s}$ and $n_{\rm CR}$ are the CR velocity and density. In the upstream region, which is not accelerated to relativistic velocities by the CR pressure, the CR velocity 
is about the speed of light, $v_{\rm s}\sim c$; thus, we have
\beq
\label{eq:Jcr}
J_{\rm CR}\sim ecn_{\rm CR} .
\eeq

To maintain overall charge neutrality, the upstream
responds to the CR current by supporting an opposite ``return
current'' ${\bf J}_{\rm ret}$ that short-circuits the electric field
associated with the charge separation in the CR precursor.\footnote{The return current is established on
the plasma time $\lesssim 10\textrm{ ms}$ which is instantaneous
compared to the light crossing time of the CR precursor.} If $n_{\rm CR}$ is smaller than the upstream density,
the return current approximately balances the CR current, 
\beq
\label{eq:Jret}
J_{\rm ret}\sim J_{\rm CR} .
\eeq  
If the
upstream contains a weak magnetic field, the return current
couples with this field; next we explore the consequences of this
coupling.

\subsection{The Driving of Turbulence and Field Amplification}
\label{sec:turbulence}

Here we argue that the return current drives turbulence in the
upstream and suggest that this turbulence leads to magnetic
field growth on scales much larger than the proton plasma skin
depth. One type of interaction between the CR precursor and the
magnetic field of the upstream, explored by
\citet{Bell:04,Bell:05} in the context of Newtonian shocks, is the
Amp\`ere force 
\beq
{\bf F}=\frac{{\bf J}_{\rm ret}\times{\bf B}}{c} 
\eeq
which accelerates the upstream perpendicular to the shock
velocity.
We assume that
the initial weak magnetic field has power on all scales, as expected
in interstellar turbulence, and that the power on 
relevant scales is roughly scale-independent.

\begin{figure}
\plotone{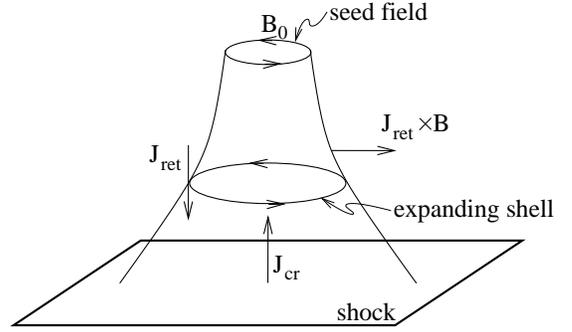}
\caption{Schematic representation of the reaction of the upstream plasma to the  CR streaming ahead of the shock.  The figure is in the frame of the shock so that the upstream plasma moves vertically downward.  An initially weak magnetic field loop is stretched by the Amp\`ere force $({\bf J}_{\rm ret}\times{\bf B})/c$ and plasma is accelerated sideways as it flows toward the shock front. \label{fig:drawing}}
\end{figure}

Consider a loop of radius $\lambda$ of a weak initial magnetic field
${\bf B}_0$ parallel to the shock transition that is
directed clockwise as seen from the shock, as shown in Figure \ref{fig:drawing}, 
and assume that the upstream is initially
stationary. The Amp\`ere  force accelerates the fluid away from the
center of the loop. \citet{Bell:04,Bell:05} carried out
magnetohydrodynamic (MHD) simulations of this process; we base our estimates on his results.

The dynamics of the upstream fluid can be approximated using MHD, and
thus flux freezing during the expansion of the loop implies
that 
\beq
\frac{B_\theta}{\rho r} =\textrm{ const},
\eeq 
where $B_\theta$  and $\rho$ are the azimuthal magnetic
field and the density at radius $r$, respectively. If the expansion is
non-relativistic and if the thermal and magnetic pressure
are ignored so that only the Amp\`ere
force influences the motion, the radius of the loop accelerates according to
\beq
\frac{d^2r}{dt^2}\sim \frac{J_{\rm ret}B_\theta}{\rho c} \sim \frac{J_{\rm CR} B_0}{\lambda \rho_0 c} r .
\eeq
Therefore, the loop expands exponentially $r(t)\sim \lambda
e^{\sigma t}$ at a rate 
\beq
\label{eq:sigma}
\sigma\sim \left(\frac{J_{\rm CR}B_0}{\lambda \rho_0 c}\right)^{1/2} ,
\eeq
implying a velocity of expansion of $dr/dt\sim r\sigma$.

The upstream accelerates until the pressure overtakes the Amp\`ere force 
(we assume that the expansion remains 
subluminal at this point). In a realistic
environment where the magnetic field fluctuates on all scales, we
expect the expansion on each
scale to saturate when neighboring magnetic ``rings'' on similar
scales collide, namely when $r\sim 2\lambda$. 
This happens when 
\beq
\label{eq:t}
\sigma t \sim \frac{\sigma\Delta}{c}\sim 1 .
\eeq
Compressional amplification of the magnetic field at this point is of the
order of unity. However it is commonly argued \citep[and references therein]{Kulsrud:05} that a turbulent dynamo
produces magnetic energy near equipartition with the turbulent energy.
Therefore, the 
upstream fluid that is exposed to a current $J_{\rm CR}$ over a
time $t$ will develop turbulent motion on all scales 
\beq
\lambda\lesssim \lambda_{\rm max} \sim \frac{J_{\rm CR} B_0 t^2}{\rho_0 c} .
\eeq

Each eddy amplifies the magnetic energy on its own scale on an 
$e$-folding time $1/\sigma(\lambda)$.
The bulk of the kinetic energy initially stored in expanding shells $\sim \onehalf(\sigma\lambda_{\rm max})^2$ turns into turbulent motion and the magnetic field on scales $\sim\lambda_{\rm max}$.  The field in equipartition with the energy in turbulent eddies equals 
\beq
B_1\sim \left(\frac{4\pi\lambda J_{\rm CR} B_0}{c\rho_0}\right)^{1/2} .
\eeq
This is the minimum field strength generated in the shock upstream on scales $\lambda\leq\lambda_{\rm max}$.
The interaction between the CRs and the generated field could continue to accelerate the fluid after the shells have collided once.  This could result in additional turbulent driving and field growth.  Then a saturation occurs when the Amp\`ere force balances the magnetic tension
\beq
|{\bf F}|\sim \frac{B^2}{4\pi\lambda} ,\eeq 
which implies a limit on the final field of 
\beq
\label{eq:B}
B_2\sim \frac{4\pi\lambda J_{\rm ret}}{c}\sim \frac{4\pi\lambda J_{\rm CR}}{c} .
\eeq 
\citet{Bell:04}, who considered Newtonian shocks, argues that such a field is generated as a consequence of CR streaming. 
Therefore we can limit the generated magnetic field to lie between $B_1$ and $B_2$.

For some $\lambda$, the above values of $B$ formally imply superluminal motion.  We do not analyze the evolution of the magnetic field after its energy approaches equipartition with the rest energy of the upstream.\footnote{Magnetic field exceeding equipartition with the rest energy of the upstream would affect the hydrodynamic profile of the shock and accelerate the upstream in the direction of shock propagation, thereby reducing $v_{\rm s}$ and $J_{\rm CR}$.} Our treatment is applicable at distances from the shock where the magnetic field does not reach equipartition, $B< (4\pi\rho_{\rm ism})^{1/2}c$.

\subsection{The Cosmic Ray Spectrum and the Quasi-Steady State}
\label{sec:spectrum}

The generated magnetic field can modify the spatial profile of 
$J_{\rm CR}$ by confining 
CRs closer to the shock than the preexisting field. The
maximum coherence length of the new field cannot exceed the
maximum distance CRs reach from the shock
$\sim R/\Gamma^2$.  Therefore
the propagation of the most energetic CRs traversing a distance $X\sim R$ between emission and re-absorption is always diffusive in the new field. If the CR motion is dominated by the new field,\footnote{The preexisting magnetic field could have power on the largest scales $\sim R$ and thus it could dominate the CR motion even if the new field is stronger.} the shock settles in a
quasi-steady state in which to every distance $\Delta\leq R/\Gamma^2$ from the shock corresponds a $\gamma$ such that CRs with Lorentz factor $\gamma$ are confined within a distance $\Delta$ from the shock by the magnetic field generated by the streaming of these same CRs.

The density of the CRs at a distance $\Delta$ is  
\beq
\label{eq:ncr}
n_{\rm CR}(\gamma)\sim \frac{1}{4\pi R^2\Delta(\gamma)} \frac{ dN}{d\ln\gamma} ,
\eeq
where $dN/d\gamma$ is the CR energy spectrum.   
The CR spectrum produced by the DSA is typically a power law 
\beq
\label{eq:dNdgamma}
\frac{dN}{d\gamma}\propto \gamma^{-p}\eeq 
in a range $\gamma_{\rm min}\leq \gamma\leq \gamma_{\rm max}$. We take the minimum Lorentz factor of CRs to equal \citep[and references
therein]{Achterberg:01}
\beq
\label{eq:gammamin}
\gamma_{\rm min}\sim \Gamma^2 .
\eeq

Numerical simulations of acceleration in ultrarelativistic shocks \citep{Achterberg:01,Ellison:02,Lemoine:03} agree with the spectral index $p=\frac{20}{9}\approx 2.22$ derived analytically by \citet{Keshet:05} for isotropic diffusion.  While the true spectral index will depend on the detailed interaction between CRs and magnetic turbulence (e.g., \citealt{Ellison:04,Lemoine:05,Niemiec:06,Lemoine:06}) and may differ from this derived value, we adopt 
\beq
\label{eq:p}
p=\frac{20}{9}
\eeq 
in what follows. If $E_{\rm CR}$ is the total energy in CRs in the shock wave, then the CR spectrum is normalized such that 
\beq
\label{eq:normalization}
\int_{\gamma_{\rm min}}^{\gamma_{\rm max}}\gamma m_{\rm p}c^2\frac{dN}{d\gamma} d\gamma = E_{\rm CR} .
\eeq

We relate the Lorentz factor of the shock to the total energy in the blast wave \citep{Blandford:76} 
\beq
\label{eq:Gamma}
\Gamma \approx \left(\frac{E_{\rm tot}}{n_0m_{\rm p}c^2R^3}\right)^{1/2} ,
\eeq 
where $n_0 \approx \rho_0/m_{\rm p}$.  We assume further that a fraction 
\beq
\label{eq:epsiloncr}
\epsilon_{\rm CR}\equiv \frac{E_{\rm CR}}{E_{\rm tot}}
\eeq
of the blast wave is stored in the CRs.  

We proceed to outline the procedure by which the  shock-accelerated CR Lorentz factor, the  shock-generated magnetic field coherence length, and the energy density in the shock generated magnetic field are calculated self-consistently as a function of distance from the shock $\Delta$.  In \S~\ref{sec:grb}, we apply this procedure to external GRB shocks.

\subsection{The Self-Consistent Solution}
\label{sec:solution}

At any distance $\Delta$ from the shock, 
the CR current is dominated by the least
energetic CRs that reach this distance. CRs with smaller $\gamma$ are confined to shorter distances from the shock. The CR current $J_{\rm CR}(\Delta)$ is larger at smaller $\Delta$.  Note the time over which an upstream element is exposed to $J_{\rm CR}(\Delta)$ is proportional to $\Delta \ll R$, 
i.e., since the cosmic rays 
are confined so close to the shock front in an ultrarelativistic shock, the time available for generation  of the magnetic field is  short.
The  field is
generated on scales $\lambda$ for which $\sigma(\lambda,\Delta) \Delta/c \gtrsim 1$..\footnote{In reality, field saturation will require multiple $e$-foldings, $\sigma \Delta/c\sim\textrm{ few}$.}   In calculating $\sigma$, the distance CRs reach from the shock and the CR number and current densities are all
calculated using the generated field $B$, rather than the preexisting field $B_0$.
A self-consistent solution for the CR and magnetic field profile ahead of the shock shows that CRs with smaller $\gamma$ generate a stronger magnetic field (because of their larger $J_{\rm CR}$) on smaller scales $\lambda$ (because of the shorter exposure time) and are confined by their self-generated field.   The maximum CR Lorentz factor $\gamma_{\rm max}$ and the maximum field correlation length $\lambda_{\rm max}$ are obtained for $\Delta \sim R/\Gamma^2$ (see \S~\ref{sec:trajectories}). 

To solve for $\gamma(\Delta)$, $\lambda(\Delta)$, and $B(\Delta)$, we proceed as follows. First, we eliminate $r_{\rm g}$ and $X$ from equations (\ref{eq:rg}), (\ref{eq:X}; the diffusive case), and (\ref{eq:Delta}) to obtain
\beq
\label{eq:one}
\lambda\Delta \sim \frac{\gamma^2 m_{\rm p}^2 c^4}{\Gamma^4 e^2 B^2}
\eeq
Next, we eliminate $n_{\rm CR}$, $\gamma_{\rm min}$, and $E_{\rm CR}$ from equations (\ref{eq:Jcr}) and (\ref{eq:ncr}---\ref{eq:epsiloncr}) to obtain
\beq
\label{eq:two}
J_{\rm CR}=\frac{\Gamma^{4/9}\epsilon_{\rm CR}E_{\rm tot}e}{18\pi\gamma^{11/9}m_{\rm p}c R^2 \Delta} .
\eeq
Next, we eliminate $\sigma$ from equations (\ref{eq:sigma}) and (\ref{eq:t}) to obtain
\beq
\label{eq:three}
\left(\frac{J_{\rm CR}B_0}{\lambda \rho_0 c}\right)^{1/2}\frac{\Delta}{c}\sim 1 .
\eeq
Finally, we set the strength of the saturated magnetic field to $B_2$  where the Amp\`ere force balances magnetic tension (see \S~\ref{sec:turbulence}; the results for saturation at $B_1$ are qualitatively the same)
\beq
\label{eq:four}
B\sim \frac{4\pi\lambda J_{\rm CR}}{c} .
\eeq

We substitute $\Gamma$ from equation (\ref{eq:Gamma}) in equations (\ref{eq:one}) and (\ref{eq:two}).  Then we substitute $J_{\rm CR}$ from equation (\ref{eq:two}) in equations (\ref{eq:three}) and (\ref{eq:four}). We finally solve for $\gamma$, $\lambda$, and $B$ as a function of $\Delta$ and the parameters of the blast wave.  It is convenient to express the magnetic field strength in terms of 
\beq
\epsilon_B\equiv \frac{B^2}{4\pi \rho c^2}
\eeq
and to express distance from the shock in terms of the dimensionless parameter
\beq
\tilde \Delta \equiv \frac{\Delta\Gamma^2}{R} \leq 1 .
\eeq
The maximum $\gamma$ and $\lambda$ correspond to CRs that reach  farthest from the shock, i.e., to $\tilde \Delta=1$.

\section{Gamma-Ray Burst Afterglows}
\label{sec:grb}

External shocks in GRBs are the best astrophysical candidates of
weakly magnetized relativistic collisionless shocks.  They have Lorentz factors $\Gamma \gtrsim 100$  when the 
blast wave is at a radius $R \sim 10^{17}\textrm{ cm}$ \citep[and references therein]{Piran:05a}. 
The blast wave decelerates and becomes Newtonian at 
$R\sim 10^{18}\textrm{ cm}$. The
external shocks are initially beamed in the form of a 
jet with a typical opening angle
$\theta_{{\rm j}} \sim 0.1\textrm{ rad}$.  The angular size of the causally connected 
region in the shock is $1/\Gamma$. At early times when $\Gamma > 1/\theta_{{\rm j}}$, the blast wave has
isotopic equivalent energy 
$E_{\rm tot}=10^{52}-10^{54}\textrm{ ergs}$ and evolves as a spherical fragment with a Lorentz factor given by equation (\ref{eq:Gamma}).
The evolution at late times when $\Gamma \lesssim 1/\theta_{{\rm j}}$ is poorly understood: the
opening angle is expected to increase; $\Gamma$ decays
faster (perhaps exponentially) with $R$, and the isotropic
equivalent energy approaches $\sim 10^{51}\textrm{ ergs}$. 

Here we explore the interaction between the CRs and
the pre-existing magnetic field in the circum-burst medium while the
evolution of the blast-wave is quasi-spherical ($\Gamma >
1/\theta_{{\rm j}}$). We expect that our results are qualitatively
applicable also when $\Gamma < 1/\theta_{{\rm j}}$ by taking 
$E_{\rm tot} \sim 10^{51}\textrm{ ergs}$ and $R \sim 10^{18}\textrm{ cm}$.  We derive the maximum Lorentz factor $\gamma$ of CRs that reach a distance $\Delta$ from the shock, the maximum magnetic field correlation length $\lambda$ at this distance, and the fraction $\epsilon_B$ of the energy density in the magnetic fields on scales $\lambda$.

Following the solution outlined in \S~\ref{sec:solution} we obtain
\bea
\gamma&\sim& 
5\times10^7\ \tilde\Delta^{0.25}\ \epsilon_{-1}^{0.62}\ B_{-6}^{0.37}\ E_{53}^{0.75}\ R_{17.5}^{-1.4}\ n_0^{-0.37},
\nonumber\\
\lambda&\sim&
10^{11}\ \tilde\Delta^{0.70}\ \epsilon_{-1}^{0.25}\ B_{-6}^{0.55}\ E_{53}^{-0.7}\ R_{17.5}^{3.04}\ n_0^{-0.55}\textrm{ cm}\nonumber\\
\epsilon_B&\sim&
10^{-2}\ \tilde\Delta^{-1.21}\ \epsilon_{-1}^{0.99}\ B_{-6}^{0.19}\ E_{53}^{1.21}\ R_{17.5}^{-3.84}\  n_0^{-1.19} ,
\eea
where $\tilde\Delta=1$ corresponds to the highest energy CRs that are accelerated in the shock.  Here, $\epsilon_{\rm CR}=0.1\epsilon_{-1}$ is the fraction of the energy $E_{\rm tot}$ carried by CRs ahead of the shock, $E_{\rm tot}=10^{53}E_{53}\textrm{ ergs}$,  $B_0=B_{-6}\ \mu\textrm{G}$, $n_{\rm ism}=n_0\textrm{ cm}^{-3}$, and $R=10^{17.5}R_{17.5}\textrm{ cm}$.  We have used $p=\frac{20}{9}$ but the dependence on $p$ is weak.
Evidently, in the shocks considered here, the field generated in the precursor has the capacity to confine CRs with higher energies than a preexisting microgauss field.\footnote{The maximum Lorentz factor to which CRs are accelerated in the pre-existing field of the ISM is $\gamma_{\rm max}(B_0)\sim 4.6\times10^6\ E_{53}^{1/2}\ B_{-6}\ n_0^{-1/2}\ R_{17.5}^{-1/2}$.}

Note that $\lambda_{\rm max}\gg \lambda_{\rm s}\sim 2\times10^7 n_0^{-1/2}\textrm{ cm}$. 
The fractional magnetic energy density $\epsilon_B$ should be approximately preserved as the fluid passes the shock transition.  The dependence of $\epsilon_B$ on the radius and the field correlation length is 
$\epsilon_B\propto R^{1.41}\lambda^{-1.73}$ 
(it is independent of $E_{\rm tot}$ if  $\epsilon_{\rm CR}$ is constant).  Note that $\epsilon_B$ is larger on smaller scales since it is generated at smaller $\Delta$ by the larger $J_{\rm CR}$ that is dominated by CRs with smaller $\gamma$.

Since the upstream is turbulent on scales $\lesssim \lambda_{\rm max}$, and since this turbulence is accompanied by density inhomogeneities, the shock transition and the fluid passing the transition are also expected to be turbulent on the same scales.  The evolution of the magnetic field after it passes the  transition is described by the laws governing MHD turbulence, and we do not attempt to describe it here. The value of $\epsilon_B$ in the downstream region will be that reached in the shock upstream, modified by any evolution that the turbulence experiences as it travels downstream past the shock transition.  

The main premise of our treatment is that the precursor is dominated by protons and thus carries a net positive current.   To compete with the protons in the distance they reach ahead of the shock, the electrons would have to have Lorentz factors $\gamma_{\rm e}>\gamma_{\rm min} m_{\rm p}/m_{\rm e}\sim \Gamma^2 m_{\rm p}/m_{\rm e}$.  Electrons with $\gamma_{\rm e}\sim 10^{8}-10^{10}$ in the rest frame of the upstream are cooled by the synchrotron mechanism on timescales much shorter than the age of the shock, and thus their acceleration to even higher energies is suppressed.  Therefore, indeed, the shock precursor at proton Lorentz factors $\gamma\gtrsim 10^{5}$ is completely dominated by protons and  carries a net positive current.

\section{Discussion}
\label{sec:discussion}

The picture presented here casts doubt on the role of TWI in collisionless shocks.  It is commonly argued that TWI generates the magnetic field that facilitates collisional coupling in the shock transition.  However, if the upstream is weakly magnetized, the shock accelerates CRs, and these CRs drive upstream turbulence and magnetic field generation. When the upstream fluid reaches the transition---where TWI is assumed to take place---the large-scale  field generated in the precursor will plausibly quench  TWI.  \citet{HededalNishikawa:05} find that TWI is quenched for ratios of the electron plasma to cyclotron frequency $\omega_{\rm pe}/\omega_{\rm ce}<5$. For nonrelativistic upstream electrons this implies quenching for $\epsilon_B\gtrsim 2\times 10^{-5}$, which is expected from our analysis.\footnote{Because of the finite transverse velocities of the CRs ($\sim c/\Gamma$ in the frame of the ISM), the CRs themselves will not excite TWI in the upstream.}

Recently, a question has been raised whether external GRB shocks are sites of UHECR acceleration 
($\gamma\sim 10^{10}-10^{11}$,~\citealt{Dermer:02,Waxman:04}). 
The idea that UHECRs are produced in GRBs \citep{Waxman:95,Vietri:95,Milgrom:95} is motivated by the observation that the cosmic energy densities in UHECRs and GRB fireballs are comparable.  
The presumed absence of strong upstream magnetic fields in external GRB shocks has focused attention on UHECR production in shocks propagating into relativistic ejecta \citep{Waxman:04,Gialis:05}.  Our analysis of external shocks indicates that the maximum Lorentz factor of CRs produced there is well below the UHECR range.  Even under the most optimistic assumptions ($\epsilon_B\sim 1$ on scales $\lambda\sim R/\Gamma^2$), the resulting Lorentz factors are
\beq
\gamma < 3\times10^{10} E_{{\rm tot},51}^{1/3} n_0^{1/6} , 
\eeq
where $E_{\rm tot}=10^{51}E_{{\rm tot},51}\textrm{ ergs}$ is the total, beaming corrected energy of the blast wave. Therefore we do not expect UHECR acceleration in external GRB shocks.

\section{Conclusions}
\label{sec:conclusions}

CRs accelerated in relativistic collisionless shocks excite large-scales turbulence and magnetic field generation in the shock upstream.  The field generated in the shock precursor has power on scales much larger than the proton plasma skin depth.  The propagation of CRs in the generated field is diffusive.  In external GRB shocks, CRs are accelerated to higher energies in the generated field than in the pre-existing field of the ISM. The generated field reaches equipartition with the energy density in the fluid.  The shock transition is turbulent with a hydrodynamic profile dominated by CR pressure. 
The commonly invoked TWI is probably quenched in relativistic collisionless shocks by the magnetic field and the turbulence generated in the shock precursor.  External GRB shocks do not accelerate UHECRs. PIC simulations of collisionless shocks must include a weak upstream magnetic field and simulate a spatial domain as large as the CR gyroradius in the upstream to observe the acceleration of particles beyond equipartition.

\acknowledgements
 
We thank Jonathan Arons and Andrei Gruzinov for inspiring us to explore the importance of upstream turbulence.  We are also indebted to Tony Bell, Don Ellison, and an anonymous referee for sending us detailed comments and to Steven Cowley and Marc Kamionkowski for illuminating conversations.  M.~M.\ thanks the Center for Cosmology and Particle Physics at New York University for hospitality and acknowledges support by NASA through the  Hubble Fellowship grant HST-HF-01188.01-A awarded by the Space Telescope Science Institute, which is operated by the Association of Universities for Research in Astronomy, Inc., for NASA under contract NAS5-26555. E.~N.\ was supported at Caltech by a senior research fellowship from the Sherman Fairchild Foundation.

\end{document}